ON THE FIGURE THAT THE WIND CAN INDUCE ON A STAGNANT FLUID

*E 494 -- De Figura Quam Ventus Fluido Stagnanti Inducere Vale*t

According to the records, it was presented to the St. Petersburg Academy on October 27, 1777.
Originally, published in *Acta Academiae Scientarum Imperialis Petropolitinae* 1777, 1778, pp. 190-194
Also in *Opera Omnia*: Series 2, Volume 13, pp. 370-375

Translated[1] and Annotated
by
Sylvio R. Bistafa[*]
May 2016

1. So long as the surface of a fluid, as well as the direction of the wind, are perfectly horizontal, there is no doubt that the fluid can persist in this state; when however this condition is disturbed by some force, it certainly can happen, that the fluid could attain another level of equilibrium with the wind. Therefore, whatever figure type these accidents to the wind should induce in the fluid by whatever chance, this is here set forth to be specified.

2. So, above the horizontal axis $OB$ let the curved figure $AYB$ be represented (Fig. 1)[2], to which the fluid is pushed by the action of the wind striking against a horizontal direction $VY$, which may remain at rest, and also be taken above the line of the vertical $AO$ the abscissa $AX = x$, to which corresponds the applied $XY = y$ and the arc $AY = s$, and point $A$ is raised to the highest surface of the fluid; where, since the fluid is supposed to be in equilibrium, its pressure at the element $Yy = ds$ will be due to the height $AX = x$, certainly the quantity of pressure $= xds$, or if

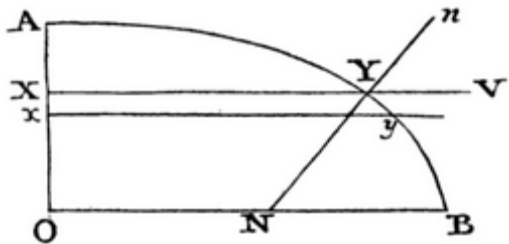
Fig. 1

leveled by a weight, which is taken as a column of the same fluid, pressed on base $ds$, whose height will be $= x$; the true direction of its pressure will be the line $Yn$ normal to the curve at the point $Y$.

3. Now set the wind velocity $= c$, so that $c$ be the distance, that the wind traverses after each second, which direction $VY$ then falls upon the element $Yy = ds$ and the angle $VYy$, whose sine is $= \frac{dx}{ds}$. But now, if $g$ indicates the height of fall of a weight in one second, that velocity of the wind will be due the height $cc/4g$[3]. Whence, if the wind were to strike the element $Yy$ directly, its force would effectively equal to the weight of a column of air, whose base $= ds$, its height however $= cc/4g$, so that if the specific gravity of the fluid to the air were just as $n$ to 1, the weight of the column, reduced to the mass of the fluid, will be $= \frac{ccds}{4ng}$ [4].

---

[1] The author gratefully acknowledges George Heine for revising this translation.
[*] Corresponding address: sbistafa@usp.br

[2] Note that the axes of the usual Cartesian coordinate system had been interchanged; here, abscissa $x$ is vertical, with its origin in $A$ (not in $O$), and the ordinate $y$ is horizontal.

[3] The generating height $s$ of which the velocity is $c$ is given by $s = c^2/2G$, where $G = 32 ft/s^2$ is gravity. A body freely falling from rest completes $16\ ft.$ in the first second; hence $g = 16\ ft$. Therefore: $s = \frac{c^2}{2G} = \frac{c^2}{4g} = \frac{c^2}{64}$.

4. Things hold in this way if the wind falls perpendicularly; on the other hand, in the case that it strikes at an angle, whose sine is $= \frac{dx}{ds}$, then according to the accepted common opinion that force must dimish in ratio to the square of the sine of the angle of incidence[5], such that the total force of the wind would be $\frac{ccds}{4ng} \cdot \frac{dx^2}{ds^2} = \frac{ccdx^2}{4ngds}$, whose direction will be the line $YN$, parallel to the normal curve, but directed to the inside; so that it is clear, that the fluid can stand still, if this force is equal to the force previously defined, applied to the exterior according to $Yn$; so on account of this matter appears this equation: $xds = \frac{ccdx^2}{4ngds}$; and it is with this equation that the nature of the curve $AYB$, which the fluid can conserve under the action of the wind, will be determined.

5. So, from here, the equation relating the abscissa $AX = x$ and the arc $AY = s$ is easily deduced, namely

$$xds^2 = \frac{ccdx^2}{4ng}; \text{ hence, } ds = \frac{cdx}{2\sqrt{ngx}};$$

and therefore, upon integrating, $s = \frac{c\sqrt{x}}{\sqrt{ng}}$, such that the arc $AY$ be proportional to the square root of the abscissa; whence it is clear that this curve will be a *cycloid*. But if we wish to introduce into the calculation the application of $XY = y$, then on account of $ds^2 = dx^2 + dy^2$ we shall have this equation: $dy = dx\sqrt{\frac{cc}{4ngx} - 1}$. Putting here $\frac{cc}{4ng} = 2a$, so that it produces $dy = dx\sqrt{\frac{2a-x}{x}}$, or $dy = dx \frac{(2a-x)}{\sqrt{2ax-xx}}$, whose integral is

$$y = \sqrt{2ax - xx} + \int \frac{adx}{\sqrt{2ax-xx}} \text{ [6]} \quad \text{or} \quad y = \sqrt{2ax - xx} + a \, Asin \frac{\sqrt{2ax-xx}}{a},$$

which is a very familiar property of the cycloid.

6. Since here the letter $a$ express the radius of the circle of this cycloid, its total height will be $= 2a$. Therefore, because we have set $2a = \frac{cc}{4ng}$, the total height of this cycloid will be $\frac{cc}{4ng}$; to regard an example, if the wind velocity be 32 feet per second, and $g$ be $= 16 \, ft.$, and for water be taken $n = 800$, the height of this cycloid will be $\frac{1}{50}$. Here it is rightly understood, that the water in this situation cannot stand, unless it be restrained with a firm wall at $AO$, so that it cannot flow.

---

[4] This corresponds to the weight of a column of fluid equivalent to the kinetic head of the wind.
[5] Euler proves this result in *Scientia navalis seu tractatus de construendis ac dirigendis navibus*, (§ 474). EULER, Leonhard. St. Petersburg, 1749.
[6] Actually, this integral is given by $aAcos\frac{a-x}{a}$.

7. However, according to experiments recently instituted, it has been discovered that the action of the fluids is to recede more in accordance with the ratio of the square of the sine of the angle of incidence, where that angle is smaller, and when the minimum angles are approached, to approximate as a simple ratio[7]; consider also this hypothesis, that the impulse of the wind along $YN$ be $= \frac{ccdx}{4ngds}$, whence it arises this equation: $xds = \frac{ccdx}{4ng}$, or $ds = \frac{ccdx}{4ngx}$ [8], so that $\frac{xds}{dx} = \frac{cc}{4ng}$; this is a constant quantity.

8. So now, with respect to the curve which is to be known, through the highest part of the fluid $A$ may be guided the horizontal $AD$, through which, from $Y$, is conducted the tangent $YT$, and from this place in the direction of $XY$ the perpendicular $TU$, and from $Y$ the perpendicular $uy$ [9], such that on that account the triangles $TYU$ and $Yyu$ will be similar; the tangent itself $YT = \frac{xds}{dx}$, which consequently, since it be equated to a constant quantity, indicates that curve to be a *tractrix* [10], which becomes visible, if in the string $TY$, whose length $= \frac{cc}{4ng}$, one extremity $T$ is conducted through the horizontal $DA$, while the other extremity $Y$ is connected to a particle, which is dragged slowly on the plane. Whence it is clear, that the highest point A is itself driven away to the left to infinity; therefore, the beginning of this curve will be $B$, where the tangent $BD$ is vertical $= \frac{cc}{4ng}$.

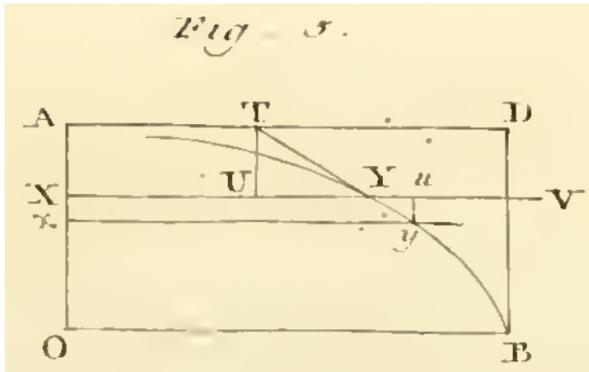

9. But if now for this curve, in place of $ds$ we write $\sqrt{dx^2 + dy^2}$, then appears the equation $x\sqrt{dx^2 + dy^2} = \frac{cc}{4ng}dx = bdx$, having set $\frac{cc}{4ng} = b$, so that $b$ exhibits the length of the pulled string, whence is deduced $dy = \frac{dx}{x}\sqrt{bb - xx}$, in which equation, having set $\sqrt{bb - xx} = z$, so that $xx = bb - zz$ and $\frac{dx}{x} = -\frac{zdz}{bb-zz}$, it transforms into this: $dy = -\frac{zzdz}{bb-zz} = dz - \frac{bbdz}{bb-zz}$; then integrating, $y = z - \frac{1}{2}bln\frac{b+z}{b-z}$; consequently,

$$y = C + \sqrt{bb - xx} - \frac{1}{2}b\,ln\frac{b+\sqrt{bb-xx}}{b-\sqrt{bb-xx}}\text{ [11]}.$$

---

[7] By applying the integral form of the momentum equation of Fluid Mechanics, it can be shown that no matter what is the angle of incidence, the action of the fluids is given by the simple ratio of the sine of the angle of incidence. See for instance: White, F. M. (1979). *Fluid Mechanics*. New York, NY: McGraw-Hill.
[8] The derivation of this result is incorrect in the original manuscript, which has been here corrected.
[9] In the original manuscript, the text describing the geometry of the accompany figure has some inconsistencies, which have been here corrected.
[10] The tractrix is perhaps less known than the cycloid and sometimes called a tractory or equitangential curve. It was first introduced by Claude Perrault in 1670, and later studied by Newton, Huygens, Leibniz, and J. Bernoulli, among others. "The tractrix is the curve along which an object moves, under the influence of friction, when pulled on a horizontal plane by a line segment attached to a tractor (pulling) point that moves at a right angle to the initial line between the object and the puller (From Wikipedia, the free encyclopedia)".
[11] Actually, the integral is given by $-\frac{1}{2}b\,ln\frac{\sqrt{bb-xx}-b}{\sqrt{bb+xx}+b}$.

Whence it exposes, supposing $= 0$, it will be that $y = C + b \ln\infty$, where is recognized the point $A$ elongated to the left at infinity.

10. Now let us combine these two hypothesis, so that we may come near enough to the truth, let the true angle of incidence be set $= \emptyset$; set the impulse of the wind to be proportional to this formula: $(1 - \alpha)\sin^2\emptyset + \alpha\sin\emptyset$, where $\alpha$ is sufficiently small. Hence, for instance, when $\emptyset$ differs little from a right angle, so that $\sin\emptyset = 1$, this formula will give 1. But if on the contrary, the angle $\emptyset$ be very small, so that $\sin^2\emptyset$ vanish in front of $\sin\emptyset$, the impulse will follow the simple calculation $\alpha\sin\emptyset$. Therefore, in our case, with $\sin\emptyset = \frac{dx}{ds}$, the force originated from the wind will be $\frac{ccds}{4ng}\left[(1-\alpha)\frac{dx^2}{ds^2} + \alpha\frac{dx}{ds}\right] = xds$ [12], from which having set $\frac{cc}{4ng} = 2a$, our equation will be $\frac{xds}{2a} = (1-\alpha)\frac{dx^2}{ds} + \alpha dx$, therefore, $\frac{xds^2}{2a} = (1-\alpha)dx^2 + \alpha dxds$, which, setting $ds = rdx$, is changed to this: $\frac{rrx}{2a} = 1 - \alpha + \alpha r$, from which is deduced $x = \frac{2a(1-\alpha+\alpha r)}{rr}$. Therefore, with $s = \int rdx = rx - \int xdr$, we will have $\int xdr = -2(1-\alpha)\frac{a}{r} - 2a\alpha \ln r$, and thus, since the abscissa $x$ defined with our variable $r$, from which likewise the arc $s$ is also expressed, being $s = 2a\alpha + \frac{4a(1-\alpha)}{r} - 2a\alpha \ln r$.

11. But if we desire the equation between $x$ and $y$, we put $dy = pdx$ and on account of $ds = dx\sqrt{1+pp}$, our equation will be $\frac{x(1+pp)}{2a} = 1 - \alpha + \alpha\sqrt{1+pp}$; whence it is deduced $x = \frac{2a(1-\alpha)}{1+pp} + \frac{2a\alpha}{\sqrt{1+pp}}$; then certainly $y = \int pdx = px - \int xdp$. Therefore, with $\int xdp = C + 2a(1-\alpha)\text{Atan }p + 2a\alpha \ln(p + \sqrt{1+pp})$, it is revealed that both coordinates $x$ and $y$ are expressed by the same third variable $p$.

12. From the preceding it is also revealed, that if $\alpha = 0$, the curve will be a Cycloid; however, if $\alpha = 1$, then a Tractrix appears, so that the curve, which we are about to obtain, possesses something between a Cycloid and a Tractrix, whose highest point will be where $p = \infty$; and the $x$ will be $= 0$ and $y = -\infty$; therefore, with respect to our curve, it will follow the characteristics of a Tractrix. Consequently, the lowest point will be found where $p = 0$, being in this case $x = 2a = \frac{cc}{4ng}$ and $y = C$. In this way, the entire curve will differ not greatly from the characteristics of a Tractrix, immediately when $\alpha$ takes a value greater than zero. It is also observed that such value can be assigned to the letter $\alpha$, even if hardly any go astray from the truth. In fact, when experiments with this objective are put into place, the value of $\alpha$ itself differs not greatly from $\frac{1}{4}$ or $\frac{1}{5}$. But although we paid no mind to motion, from here it is not at all obscurely understood, the manner in which water can be raised to a higher level by the wind. Nevertheless it seems hardly to be hoped, that such a Theory as the above can be advanced to the motion of fluid.

---

[12] The first square bracket is misplaced in the original manuscript.